\documentclass[twocolumn,preprintnumbers,amssymb,amsmath,superscriptaddress,letterpaper,nofootinbib]{revtex4}[12pt]
\usepackage{times,graphics}
\usepackage[colorlinks=true,allcolors=blue]{hyperref}

\usepackage{graphicx}
\usepackage{dcolumn}
\usepackage{bm}
\usepackage{natbib}
\usepackage{pstricks}
\usepackage{epsfig}
\usepackage{epstopdf}
\usepackage{multirow}
\usepackage{amsmath}
\usepackage{lipsum}

\newcommand{\be}{\begin{equation}}
\newcommand{\ee}{\end{equation}}
\newcommand{\bea}{\begin{eqnarray}}
\newcommand{\eea}{\end{eqnarray}}

\makeatletter
\newcommand*{\shifttext}[2]{%
	\settowidth{\@tempdima}{#2}%
	\makebox[\@tempdima]{\hspace*{#1}#2}%
}
\makeatother

\begin{document}

\title{Cosmic voids are emptier in the presence of symmetron's domain walls}

\author{Bahar Nosrati}
\email{b.nosrati@sbu.ac.ir}
\affiliation{Department of Physics, Shahid Beheshti University, G.C., Evin, Tehran 19839, Iran}

\author{Nima Khosravi}
\email{n-khosravi@sbu.ac.ir}
\affiliation{Department of Physics, Shahid Beheshti University, G.C., Evin, Tehran 19839, Iran}

\date{\today}

\begin{abstract}
The symmetron field has an environment (density) dependent behavior which is a common feature of the models with the screening mechanism and results in a rich phenomenology. This model can produce domain walls between regions with different densities. We consider this aspect and study the physics of domain walls in between (underdensity) voids and (overdensity) halo structures. The (spherical) domain walls exert a repulsive force on a test mass outside of the wall while a test mass inside of the wall sees no force. This makes the structures outside the voids go further to a larger radius. Effectively, this means the voids are becoming larger in this scenario in comparison to the standard model of cosmology. Interestingly, this makes voids emptier which may shed light on Peebles' void phenomenon.

\end{abstract}
\maketitle

\section{Introduction}

The field of cosmology is in one of its interesting milestones. The standard model of cosmology, $\Lambda$CDM, is very successful in describing many of the different observations including the cosmic microwave background, the large structure formations, etc. But meanwhile, some of the new observation hint at anomalies and tensions in $\Lambda$CDM. These anomalies can be categorized roughly to i) discrepancies between the observations of the early universe and the late one e.g. the Hubble tension \cite{Riess:2019cxk,Bernal:2016gxb,Riess:2018byc}, $\sigma_8$ tension \cite{Heymans:2020gsg} and CMB lensing \cite{Planck:2018vyg}, ii) some anomalies are spatial anomalies e.g. the CMB dipole modulations \cite{Eriksen:2003db,Hansen:2008ym} and quadrupole-octopole alignment \cite{Copi:2006tu,Copi:2005ff,Copi:2010na,Copi:2013jna,deOliveira-Costa:2003utu,Schwarz:2004gk} and iii) the third category is in the physics of (nonlinear) cosmological structures e.g. the core-cusp problem \cite{deBlok:2009sp,Flores:1994gz} and the void phenomenon \cite{Peebles:2001nv}. We do not claim these different tensions are not related and there are proposals in the literature to study some of them altogether \cite{Banihashemi:2018has}. On the other hand, there are some theoretical issues in $\Lambda$CDM: the cosmological constant problem and the physics behind both dark matter and dark energy. The current situation convinced the community to explore different paths both in the observations and theoretical models to make our understanding deeper.

Among all of the observational probes, the cosmological voids are the less studied ones but they have become more interesting recently \cite{Pisani:2019cvo}. The physics (and evolution) of voids is correlated to the overdensity structures, both are given by one model e.g. $\Lambda$CDM model. Void physics is, however, linear, while (overdensed) structures are nonlinear. This fact makes voids an interesting cosmological probe in addition to the large scale (overdense) structures and CMB.  Though there are few tensions concerning voids when we assume $\Lambda$CDM. The void phenomenon mentioned by Peebles \cite{Peebles:2001nv} says that the observed voids are emptier than the prediction of $\Lambda$CDM. This fact can be interpreted as less structure in the voids and may hint at an environment dependent structure formation physics \cite{Peebles:2001nv}. There is another issue concerning voids: Tavasoli reported that the structures in voids are more concentrated in the center of them which is not consistent with $\Lambda$CDM's prediction \cite{Tavasoli:2021reo}. Is there any explanation for these observations?

In the theory of $\Lambda$CDM, the physics of dark energy was/is always a matter of question. For many different reasons, the modified gravity models \cite{Joyce:2014kja, Amendola:2016saw,Clifton:2011jh} are suggested instead of just having a constant, the cosmological constant. In this work, for our purposes, we focus on the symmetron model \cite{Hinterbichler:2011ca}. This model suggests that the gravitational force behaves differently according to the density. The change in its behavior is realized in a way very similar to the phase transition. When a system experiences phase transition, it has undergone symmetry breaking and in a model such as symmetron model $\rm Z_2$ symmetry, a discrete symmetry, breaks, and as a consequence domain walls (DWs) are created. This is also a natural prediction in the GLTofDE model \cite{Banihashemi:2018has}. One should be careful that these DWs are produced in very small energy scales at late times which can satisfy the observational constrains\footnote{Note that the DWs which are produced by the electroweak phase transition at very high energy scales cannot satisfy the observational tests. It is one of the main reasons for cosmological inflation \cite{Mukhanov:2005sc}.}. The symmetron DWs should be placed in the region between high and low densities. Consequently, we expect to have them at the border between the cosmological overdensities (nonlinear structures) and underdensities (voids).

In this paper, we study the effects of such DWs and their effects on the (size) distribution of cosmic voids. In the next section, we briefly review the symmetron mechanism and the production of DWs in it. Then we study the behavior of a spherical DW and especially consider the movement of a test mass in the presence of a wall. We close our paper with conclusions and future perspectives.

\section{Symmetron Model}\label{SM}
As it was discussed in the previous section one of the models which can produce DWs due to symmetry breaking is the symmetron model. We will introduce some of its parameters in this section. Symmetron action was presented by \cite{Hinterbichler:2011ca} as
\begin{eqnarray}
\nonumber
	S&=&\int d^4x\,\sqrt{-g}\bigg[\frac{M^2_{Pl}}{2}\,R-\frac{1}{2}\,g^{\mu \nu}\,\partial_{\mu}\phi\,\partial_{\nu}\phi-V(\phi)\bigg]
	\\
	&+&\int d^4x \sqrt{-\tilde{g}}\mathcal{L}_m (\psi,\tilde{g}_{\mu \nu})\,,
\end{eqnarray}
where $\tilde{g}_{\mu\nu}\equiv A^2(\phi)\, g_{\mu\nu}$. The conformal coupling factor is
\begin{eqnarray}
	A^2(\phi)&=&1+\frac{\phi^2}{2\,M^2}\,,
\end{eqnarray}
and the potential has the following form
\begin{eqnarray}
	V(\phi)&=&-\frac {1} {2} \,\mu^2\,\phi^2 + \frac {1} {4} \lambda \,\phi^4\,.
\end{eqnarray}
We can introduced the effective potential as \cite{Hinterbichler:2011ca}
\begin{eqnarray}
	V_{eff}(\phi)=\frac {1} {2} \, (\frac{\rho}{M^2}-\mu^2)  \, \phi^2 + \frac {1} {4}\, \lambda\, \phi^4\,.
\end{eqnarray}
Vacuum expectation value before the symmetry breaking would be $\phi_0=0$ and after the symmetry breaking would be:
\begin{eqnarray}\label{phi}
	\phi_0&=&\pm\frac{\mu}{\sqrt{\lambda}}\,.
\end{eqnarray}
Also \cite{Llinares:2014zxa} has introduced another form for the effective potential;
\begin{eqnarray}\label{Veff-pot}
	V_{eff}(\phi)&=&\frac {1} {4\lambda_0^2} \, \Big(\frac{\rho}{\rho_{SSB}}-1\Big)  \, \phi^2 + \frac {1} {4} \lambda \,\phi^4\,.
\end{eqnarray}
where $\lambda_0=\frac{1}{\sqrt{2}\, \mu}$ is the Compton wavelength, and $\rho_{SSB}=\mu^2\, M^2$ is the symmetry breaking density. We assume the field becomes tachyonic around the symmetry breaking redshift, $z_{SSB}$ i.e.   $ \mu^2\,M^2 \sim \alpha\, M_{pl}^2\,H_{SSB}^2$ where $H_{SSB}$ is the Hubble parameter at $z_{SSB}$ , $M_{pl}$ is the Planck mass and $\alpha$ is a free parameter. This assumption seems  natural since it means the symmetry breaking (roughly) depends only on the local (environmental)  temporal and spatial parameters. We will see that different values of the free parameter $\alpha$ plays an important role in the results by giving larger values for the DW's surface tension (which plays a key role in the effect of DW on a test particle).

So far we have presented the important parameters of the symmetron model. Since the symmetron model is a screening mechanism, there are strong constraints on its effect by local tests of gravity. Here are some constraints set by \cite{Hinterbichler:2010es}
\begin{eqnarray}
	M &\lesssim& 10^{-3} M_{pl}\,,
	\\
\label{beta}
	\beta=\frac{\phi_0 M_{pl}}{M^2} &\approx& 1\,,
	\\
	\lambda_0 &\lesssim& 10^{-3}\, c\, \alpha^{-1}\, H_{0}^{-1} \,(1+z_{SSB})^{-3/2}\,,
\end{eqnarray}
where $\beta$ is a dimensionless coefficient. Note that this is only valid in a matter dominated universe which is the regime of our interests. From relations (\ref{phi}), (\ref{beta}), and our assumption ($ \mu^2\,M^2 \sim M_{pl}^2\,H_{SSB}^2$), one is lead to
\begin{eqnarray} \label{VEV}
	\phi_0&=&6\,\beta\, \lambda_0^2\, M_{pl}\,\alpha^2 H_{0}^2\, \Omega_0\,(1+z_{SSB})^{3}.
\end{eqnarray}
where $\Omega_0$ is the background density parameter. It is evident from (\ref{Veff-pot}) that the symmetron's potential undergoes a symmetry breaking in under-dense regions, $\rho<\rho_{SSB}$. Breaking of symmetron's discrete symmetry will result in the existence of DWs. In \cite{Llinares:2014zxa} it is suggested that the surface energy density of a DW is the difference between the energy density of a universe with a DW and without one,
\begin{eqnarray}
	\sigma=\frac{2\,\phi_0^2}{3\,c\,\hbar\,\lambda_0}\,\sqrt{1-\frac{(1+z)^{3}}{(1+z_{SSB})^{3}}}\,\bigg[1+\frac{(1+z)^{3}}{2(1+z_{SSB})^{3}}\bigg].
\end{eqnarray}

\begin{figure}[h!]
	\centering
	\includegraphics[width=0.45\textwidth]{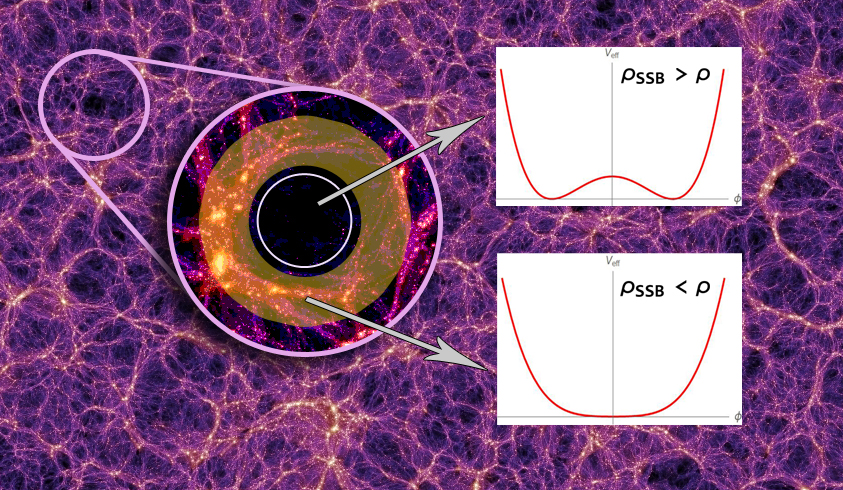}
	\caption{In this figure, we illustrate our idea with a cartoon. The symmetron field changes its behavior with respect to the environment density. In high densities (i.e. structures which are shown in a yellow-ish stripe), $\rho_{SSB}<\rho$, the effective potential (\ref{Veff-pot}) has a minimum at $\phi_0=0$ as it is shown at the lower-right plot. But in low densities (i.e. voids which are dark in the simulation), $\rho_{SSB}>\rho$, the potential experiences a symmetry breaking resulting in equilibrium at $\phi_0\neq 0$ (shown at the upper-right plot). These different values of the field cause a DW between the structures and voids (shown as the white circle). It is obvious that in the real world we cannot assume a spherical DW but in this work we keep this approximation \cite{Springel:2005nw}.}
	\label{fig.DW.Void}
\end{figure}

Observationally, we expect to have these DWs at the boundary of cosmological voids and (overdensity) structures as shown in figure \ref{fig.DW.Void}. Since in voids, we can have the negative value for the first term in (\ref{Veff-pot}) while it is positive for the structures. This means the field does (not) see the symmetry breaking in voids (structures). Consequently, the region between voids and structures has potential to host a DW, To study the effects of such DWs on the void environment, we study the geometrical properties of DWs in the next section.

\section{Spherical domain wall's metric}\label{FM}
For our purposes, we focus on the spherical DWs, they are able to give us a clue about what we should expect in observation. In the first step, we find the metric for a moving spherical DW and then we switch to the static ones.

\subsubsection{Moving Spherical Domain Wall}
In order to find the metric of spherical DW, we have considered the general form of spherically symmetric metric
\begin{eqnarray}\label{met-gen}
\nonumber
	ds^2|_+&=&e^{A(r,t)}\,dt^2-e^{B(r,t)}\,dr^2-r^2\,d\Omega^2\,,
	\\
	ds^2|_-&=&e^{D(r,t)}\,dt^2-e^{C(r,t)}\,dr^2-r^2\,d\Omega^2\,,
\end{eqnarray}
where the signs $+$ and $-$ refer to the outside and inside of the DW. 

As a starting point for solving Einstein's equations, we have assumed the following constraints: First, we are in the thin-wall limit and the wall is moving, which means; it can either collapse or expand. Moreover, the wall is assumed to be in a vacuum, and the outside metric is asymptotically flat at infinite distances. Solutions to the Einstein equations yield the metric
\begin{eqnarray} 
\nonumber
	ds^2|_+&=&e^{A(t)}\,\big(1+\frac{F}{r}\big)\,dt^2-(1+\frac{F}{r})^{-1}\,dr^2-r^2\,d\Omega^2\,,
	\\
\nonumber
	ds^2|_-&=&e^{D(t)}\,\big(1+\frac{H}{r}\big)\,dt^2-(1+\frac{H}{r})^{-1}\,dr^2-r^2\,d\Omega^2\,.
\end{eqnarray}
Here $F$ and, $H$ are the constant of integration. Since we should prevent the inside metric from becoming singular, $H$ should be zero. Consequently; the metric  will have the following form:
\begin{eqnarray} 
	\nonumber
	\label{Met_time}
	ds^2|_+&=&e^{A(t)}\,\big(1+\frac{F}{r}\big)\,dt^2-(1+\frac{F}{r})^{-1}\,dr^2-r^2\,d\Omega^2\,,
	\\
	ds^2|_-&=&e^{D(t)}\,dt^2-dr^2-r^2\,d\Omega^2\,.
\end{eqnarray}
To find $F$, $A(t)$, and $D(t)$, one should satisfy the two Israel junction conditions. The first junction condition implies that the intrinsic three-metric, $h_{ab}$, for the hypersurface (in this case the DW) be the same both inside and outside of the hypersurface ($h_{ab}|_+=h_{ab}|_-$). The second condition that was defined by Israel \cite{Israel:1966rt} is the equality of the following equations
\begin{eqnarray}
\label{eq.gamma.K}
	\gamma_{ab}&=&K_{ab}|_+-K_{ab}|_-\,,
	\\
\label{eq.gamma.S}
	\gamma_{ab}&=&8\pi\, G\,\big(S_{ab}-\frac{1}{2}h_{ab}\,S_{c}^c\big)\,,
\end{eqnarray}
where $K_{ab}$ is the extrinsic curvature and, $S_{ab}$, the stress-energy tensor, is
\begin{eqnarray}\label{eq.S}
	S^{ab}&=&\sigma u^a u^b-\tau(h^{ab}+u^a u^b)\,.
\end{eqnarray}
$\sigma$ and, $\tau$ are respectively the surface energy density and tension of the wall measured by any observer whose world line lies on the hypersurface (DW) and sees no energy flux in her local frame, and $u^a$ is the four-velocity of that observer. For the choice of equation (\ref{eq.S}), equation (\ref{eq.gamma.S}) becomes
\begin{eqnarray}
	\gamma_{ab}&=&-4\,\pi\,G \big[\sigma\, h_{ab}+2\,(\sigma-\tau)\,u_a\,u_b\big]\,.
\end{eqnarray}
For DWs, $\sigma=\tau$, implying that
\begin{eqnarray}
	\gamma_{ab}&=&-4\,\pi \,G\,\sigma \,h_{ab}\,.
\label{eq.gamma.h}
\end{eqnarray}
As a result, for a DW to satisfy the second junction condition, we must equalize (\ref{eq.gamma.K}) and (\ref{eq.gamma.h}).

To determine the intrinsic metric of the  DW and its extrinsic curvature, we should calculate the unit spacelike normal for both metrics.
\begin{eqnarray}
\nonumber
	n_{\mu}|_+&=&\Big(-e^{\frac{A(t)}{2}}\Dot{R},\,\big(1+\frac{F}{r}\big)^{-1}\sqrt{1+\frac{F}{r}+\Dot{R}^2},\,0,\,0\Big)\,,
	\\
	n_{\mu}|_-&=&\Big(-e^{\frac{D(t)}{2}}\,\Dot{R},\sqrt{1+\Dot{R}^2},\,0,\,0\Big)\,,
\end{eqnarray}
where $R$ is the radius of the DW. Be aware that dot denotes a derivative with respect to proper time and prime denotes a derivative with respect to $r$.

Now we can calculate the intrinsic metric of the DW using the outside and inside metric
\begin{eqnarray}
	\nonumber
	h_{a b}|_+&=&diag\Big(e^{A(t)}\,\big(1+\frac{F}{R}-\Dot{R}^2\big),\,-R^2,\,-R^2\sin^2{\theta}\Big)\,,
	\\
	h_{ab}|_-&=&diag\Big(e^{D(t)}(1-\Dot{R}^2),\,-R^2,\,-R^2\sin^2{\theta}\Big)\,.
\end{eqnarray}
The first junction condition indicates that $h_{a b}|_+=h_{a b}|_-$, therefore this will lead us to
\begin{eqnarray}\label{MSDW-1st}
	\frac{e^{D(t)}}{e^{A(t)}}&=&\frac{1+\frac{F}{R}-\Dot{R}^2}{1+\Dot{R}^2}\,.
\end{eqnarray}
Here we have found a relation between the metrics coefficients.

In order to satisfy the second junction condition, we need to find the extrinsic curvature for both metrics from $K_{ab}=h_a^c \bigtriangledown_c n_b$. The extrinsic curvature for the outside metric has the following form
\begin{eqnarray}
\nonumber
	K_{tt}|_+&=&-e^{\frac{-A(t)}{2}}\,\Ddot{R}-\frac{F}{2R^2}\,e^{A(t)}\sqrt{1+\frac{F}{R}+\Dot{R}^2}\,,
	\\
	K_{\theta\theta}|_+&=&R\,\sqrt{1+\frac{F}{R}+\Dot{R}^2}\,,
	\\
\nonumber
	K_{\phi\phi}|_+&=&R\,\sin^2{\theta}\,\sqrt{1+\frac{F}{R}+\Dot{R}^2}\,,
\end{eqnarray}
and for the inside metric is
\begin{eqnarray}
\nonumber
	K_{tt}|_-&=&-e^{\frac{D(t)}{2}}\,\Ddot{R}\,,
	\\
	K_{\theta\theta}|_-&=&R\,\sqrt{1+\Dot{R}^2}\,,
	\\
\nonumber
	K_{\phi\phi}|_-&=&R\,\sin^2{\theta}\,\sqrt{1+\Dot{R}^2}\,.
\end{eqnarray}
Hence to compare the extrinsic curvature with stress-energy tensor relation one finds $\gamma_{ab}$ from (\ref{eq.gamma.K}),
\begin{eqnarray}\label{gamma-1}
\nonumber
	\gamma_{tt}&=&-e^{\frac{-A(t)}{2}}\,\Ddot{R}-\frac{F}{R^2}\,e^{A(t)}\sqrt{1+\frac{F}{R}+\Dot{R}^2}+e^{\frac{D(t)}{2}}\Ddot{R}\,,
	\\
	\gamma_{\theta\theta}&=&R\,\Big(\sqrt{1+\frac{F}{R}+\Dot{R}^2}-\sqrt{1+\Dot{R}^2}\,\Big)\,,
	\\
\nonumber
	\gamma_{\phi\phi}&=&R\,\sin^2\theta\,\Big(\sqrt{1+\frac{F}{R}+\Dot{R}^2}-\sqrt{1+\Dot{R}^2}\,\Big)\,.
\end{eqnarray}
$\gamma_{ab}$ is also given by Eq. (\ref{eq.gamma.h})
\begin{eqnarray}\label{gamma-2}
\nonumber
	\gamma_{tt}&=&-4\,\pi \,G\,\sigma\, e^{D(t)}(1-\Dot{R}^2)\,,
	\\
	\gamma_{\theta\theta}&=&4\,\pi\, G\,\sigma\, R^2\,,
	\\
\nonumber
	\gamma_{\phi\phi}&=&4\,\pi\, G\,\sigma\, R^2\,\sin^2\theta\,.
\end{eqnarray}
From equalizing $\gamma_{tt}$ ((\ref{gamma-1}) and (\ref{gamma-2})) we will obtain
\begin{eqnarray} \label{2nd.MSDW.tt}
\nonumber
	-e^{\frac{-A(t)}{2}}\,\Ddot{R}&-&\frac{F}{R^2}\,e^{A(t)}\,\sqrt{1+\frac{F}{R}+\Dot{R}^2}+e^{\frac{D(t)}{2}}\,\Ddot{R} 
	\\
	&=&-4\,\pi\, G\,\sigma\, e^{D(t)}\,(1-\Dot{R}^2)\,,
\end{eqnarray}
and from $\gamma_{\theta\theta}$ (or $\gamma_{\phi\phi}$) we will have
\begin{eqnarray} \label{2nd.MSDW.thth}
	F=R\,\big(4\pi\,G\,\sigma\,R+\sqrt{1+\Dot{R}^2}\,\big)^2-(1+\Dot{R}^2)\,R\,,
\end{eqnarray}
which they have satisfied the second junction condition.

We are curious to obtain these equations for a static wall as well. To solve the first equation (\ref{2nd.MSDW.tt}) for a static wall, one should put all $\dot{R}$ to zero, but in this equation, the $\ddot{R}$ will remain which tells us that static DWs are unstable. If we convert (\ref{2nd.MSDW.thth}) to SI units, this part $1+(\frac{\Dot{R}}{c})^2$ will tell us that the speed of the DW does not play a key role in this equation. We will see in the next section that $1+(\frac{\Dot{R}}{c})^2$ can be used as a correction amplitude for a static wall solution. 

Though we have mentioned that static walls are unstable, for sake of simplicity we are going to ignore the tiny effects of DW's movement and find the metric of static DW.

\subsubsection{Static Spherical Domain Wall}
For a static wall, we can put $\dot{R}=0$ and re-write the above equations. You can find detailed calculations in the Appendix. So if the metric of the wall is
\begin{eqnarray} \label{metric-static}
\nonumber
	ds^2|_+&=&e^{A}\,\big(1+\frac{F}{r}\big)\,dt^2-(1+\frac{F}{r})^{-1}\,dr^2-r^2\,d\Omega^2\,,
	\\
	ds^2|_-&=&e^{D}\,dt^2-dr^2-r^2\,d\Omega^2\,.
\end{eqnarray}
Since wall is static, we have $A(t)=A$ and $D(t)=D$. We will get the following relation between the parameters due to the junction conditions	
\begin{eqnarray}
\label{1st.SSDW}
	\frac{e^{D}}{e^{A}}&=&1+\frac{F}{R}\,,
	\\
\label{2nd.SSDW.thth}
	F&=&R\,\big(4\pi\,G\,\sigma\,R+1\,\big)^2-R\,.
\end{eqnarray}
As it was explained for the moving DW, these equations explain the relation between the metric's coefficients and give us $F$. It is clear that equation (\ref{2nd.MSDW.thth}) will be reduced to (\ref{2nd.SSDW.thth}) when $\dot{R}$ is zero or even when the wall is moving slowly since the sentence $1+(\frac{\Dot{R}}{c})^2$ was presumed to be a correction amplitude. Therefore we conclude that it is possible to neglect the movement of the wall. Here we have not written the counterpart of equation (\ref{2nd.MSDW.tt}) for static DW, since it cannot be trusted due to having $\ddot{R}$. We will explain in the next section that ignoring the instability of static walls will not cause us any problems.

\section{Geodesic of a test particle near a static domain wall}\label{Geo}
Now we want to find the effects of the static wall on a test mass. Since solving geodesic equations for (\ref{metric-static}) is difficult we will use a shortcut and solve $u^{\mu}\,u_{\mu}=+1$ using the Killing vectors
\begin{eqnarray}
	\nonumber
	K_{\mu}&=&\Big(e^{A}\,(1+\frac{F}{r}),\,0,\,0,\,0\Big)\,,
	\\
	R_{\mu}&=&\Big(0,\,0,\,0,-r^2\,\sin^2\theta\Big)\,.
\end{eqnarray}
Hence by solving the four-velocity equation,
\begin{eqnarray}
\nonumber
	&e^{A}&\,(1+\frac{F}{r})\,(\frac{dt}{d\tau})^2-(1+\frac{F}{r})^{-1}\,(\frac{dr}{d\tau})^2
	\\
	&-&r^2\,(\frac{d\theta}{d\tau})^2-r^2\,\sin^2\theta \,(\frac{d\phi}{d\tau})^2 =+1\,,
\end{eqnarray}
neglecting the test particle's movement in $(\theta,\phi)$ plane due to symmetries, and using the following conserved quantities,
\begin{eqnarray}
\nonumber
	E&=&K_{\mu}\,\frac{dx^{\mu}}{d\lambda}=e^{A}\,(1+\frac{F}{r})\,\frac{dt}{d\tau}\,,
	\\
	L&=&-R_{\mu}\,\frac{dx^{\mu}}{d\lambda}=r^2\,\sin^2\theta\,\frac{d\phi}{d\tau}\,,
\end{eqnarray}
we will have the radial equation as
\begin{eqnarray} \label{geo-radial}
	\frac{d^2r}{d\tau^2}&=&\frac{F}{2r^2}\,.
\end{eqnarray}
Since from (\ref{2nd.SSDW.thth}) we can see that $F$ is positive, it is without question that a test particle near a spherical DW will experience repulsion. From equations (\ref{2nd.SSDW.thth}) and (\ref{geo-radial}) we understand that the radial movement of the test particle is only influenced by the surface energy density and radius of the DW.

Before showing our results we must arise the issue of the instability of static DWs. In \cite{Llinares:2014zxa}, it has been stated that when spherical DWs are pinned to matter, they become stable. Therefore by trusting their results we will examine the effect of a static spherical DW on a test particle.

\begin{figure}[h!]
	\centering
	\includegraphics[width=0.45\textwidth]{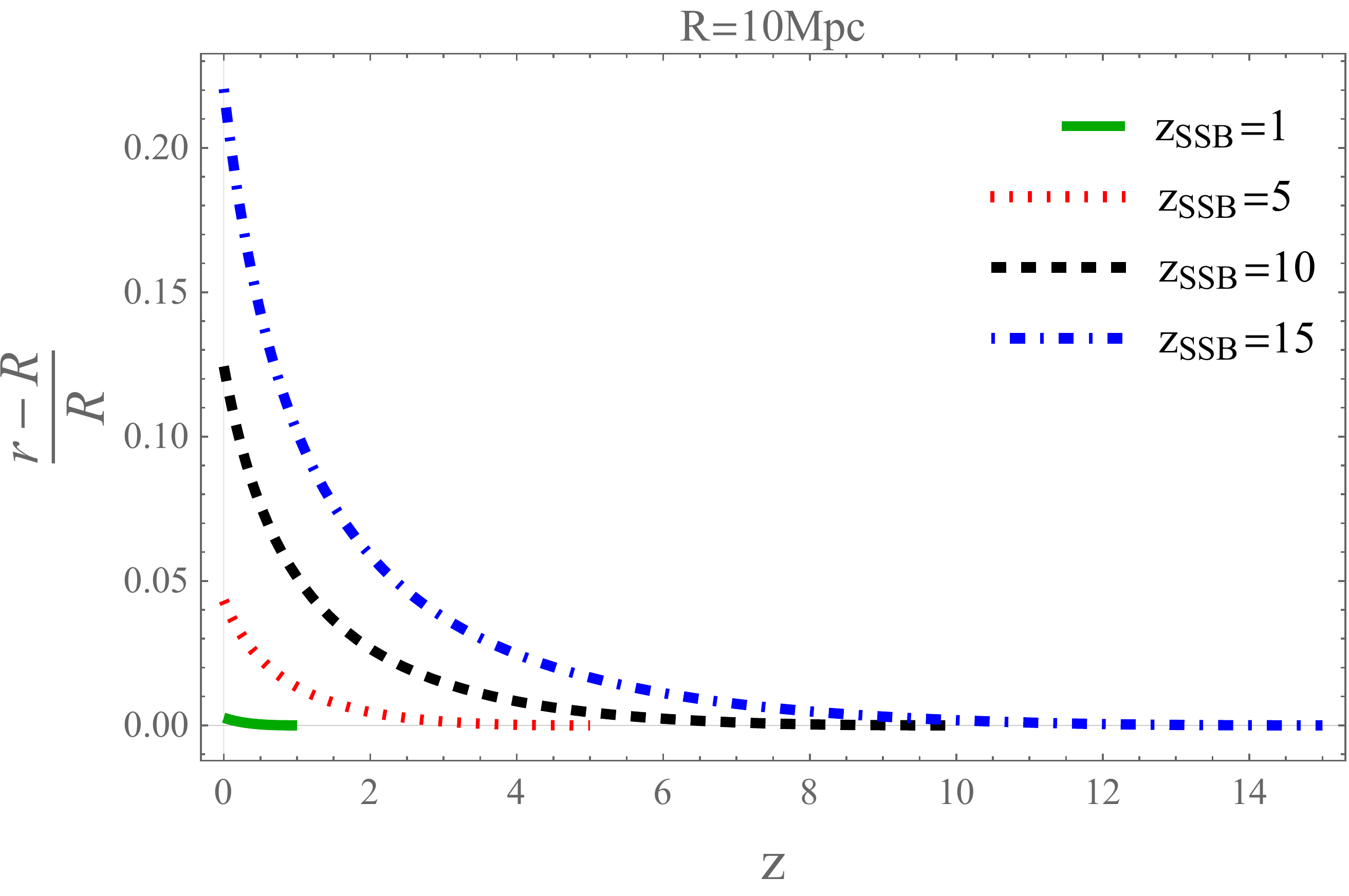}
	\caption{For a typical initial size of a void, $10 Mpc$, we have plotted the change in its radius as a function of redshift (where we fixed $\alpha=30$). As we could expect, earlier symmetry breaking causes more effect i.e. larger changes in the void size.}
	\label{fig.R10-100_z1-15}
\end{figure}

\begin{figure}[h!]
	\centering
	\includegraphics[width=0.45\textwidth]{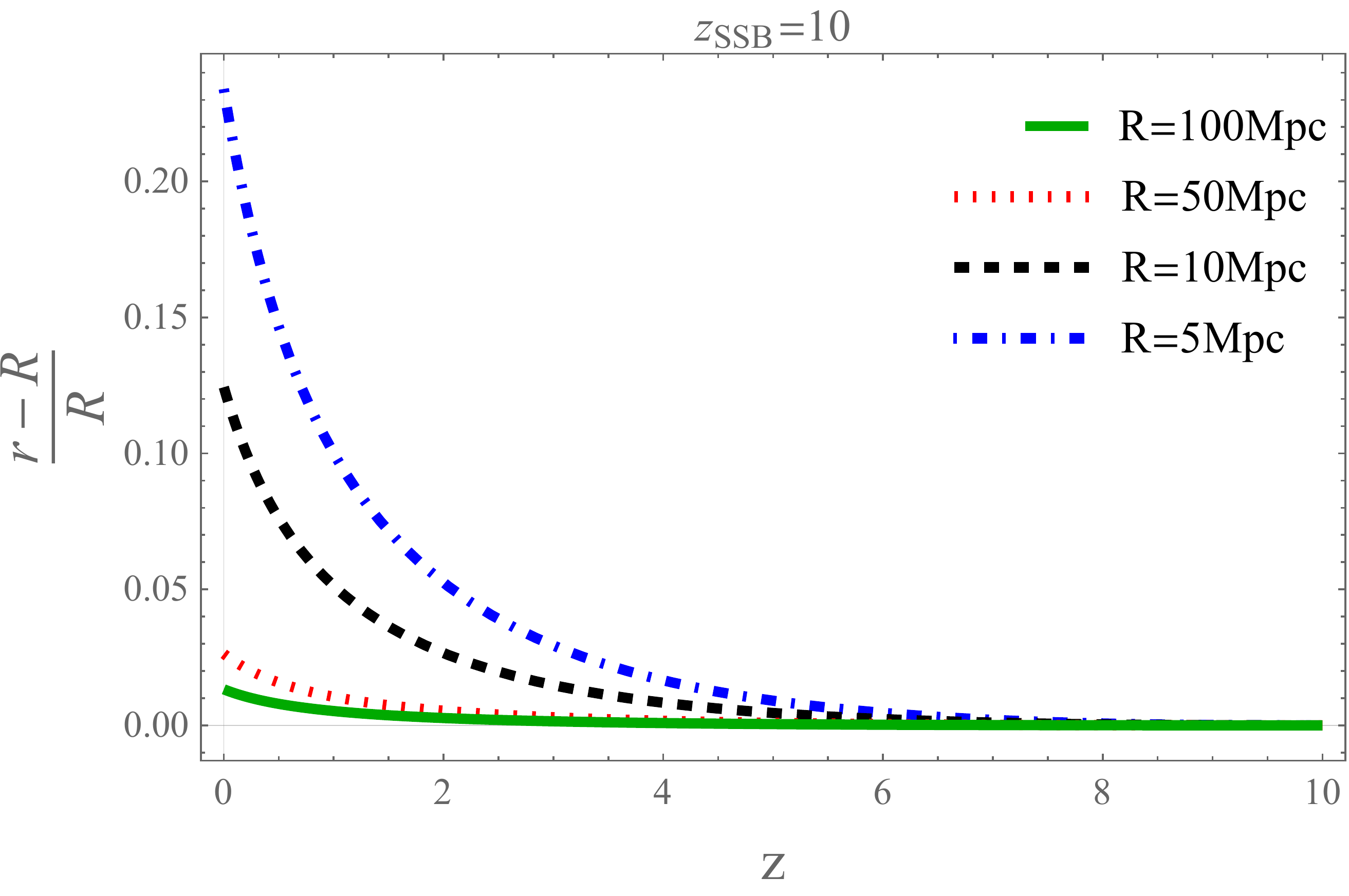}
	\caption{For a fixed $z_{SSB}=10$ and $\alpha=30$ we have plotted the change in the void's radius as a function of redshift for the different initial sizes of voids. Interestingly, we see that the relative change is larger for the smaller voids.}
	\label{fig.z1-10-R5-100}
\end{figure}

To find the effect of symmetron model DWs, we should change equation (\ref{geo-radial}) so that it will be connected to cosmological parameters. First, One should change proper time $\tau$ to cosmological time via $d\tau^2=g_{00}\,dt^2$, and $t$ should be converted into redshift. Therefore, equation (\ref{geo-radial}) becomes
\begin{equation} \label{ddotr}
	\frac{d^2r}{dz^2}
	=\frac{F}{2r^2}\,\big(1+\frac{F}{r}\big)\,\frac{1}{H_0^2\,\Omega_0 \,(1+z)^{3}}
	\,.
\end{equation}
for a matter dominant universe. We have solved equation (\ref{ddotr}) for a test particle near a spherical DW, with the following parameters
\begin{eqnarray}
\nonumber
	H_0 &\approx& 10^{-10}\, \frac{1}{yr}\,,
	\\
\nonumber
	M_{pl} &\approx& 10^{-8}\, kg\,,
	\\
\nonumber
	G &\approx& 10^{-64}\,\frac{Mpc^3}{kg \,yr^2}\,,
	\\
\nonumber
	\hbar &\approx& 10^{-72} \,\frac{Mpc^2 \,kg}{yr}\,,
	\\
\nonumber
	\Omega_0&=&1 \,,
	\\
\nonumber
	z_0&=&0 \,.
\end{eqnarray}
In figure \ref{fig.R10-100_z1-15}, we have shown the changes in void's radius as a function of redshift for a typical void with an initial radius $10 Mpc$ and free parameter $\alpha=30$. As we expected for earlier $z_{SSB}$ we see a larger effect. It is worth mentioning that the effect of DW can make a typical void with radius $10 Mpc$ to be larger by $\sim 10\%$  depending on the symmetry breaking redshifts. In figure \ref{fig.z1-10-R5-100}, we have the same plot but for different initial sizes for voids while $z_{SSB}=10$ and free parameter $\alpha=30$. It is obvious from this plot that for smaller voids, the effect is larger. This can be confirmed, mathematically, by rewriting equation (\ref{ddotr}) as
\begin{eqnarray}\label{ddotr-2}
	\frac{d^2x}{dz^2}&\propto&\frac{1}{R^2\,x^2}\,
\end{eqnarray}
where we have defined $x\equiv r/R$  and omit non-relevant terms. This means for larger $R$'s we have less changes in the $x$ which is the normalized displacement.

\begin{figure}[h!]
	\centering
	\includegraphics[width=0.45\textwidth]{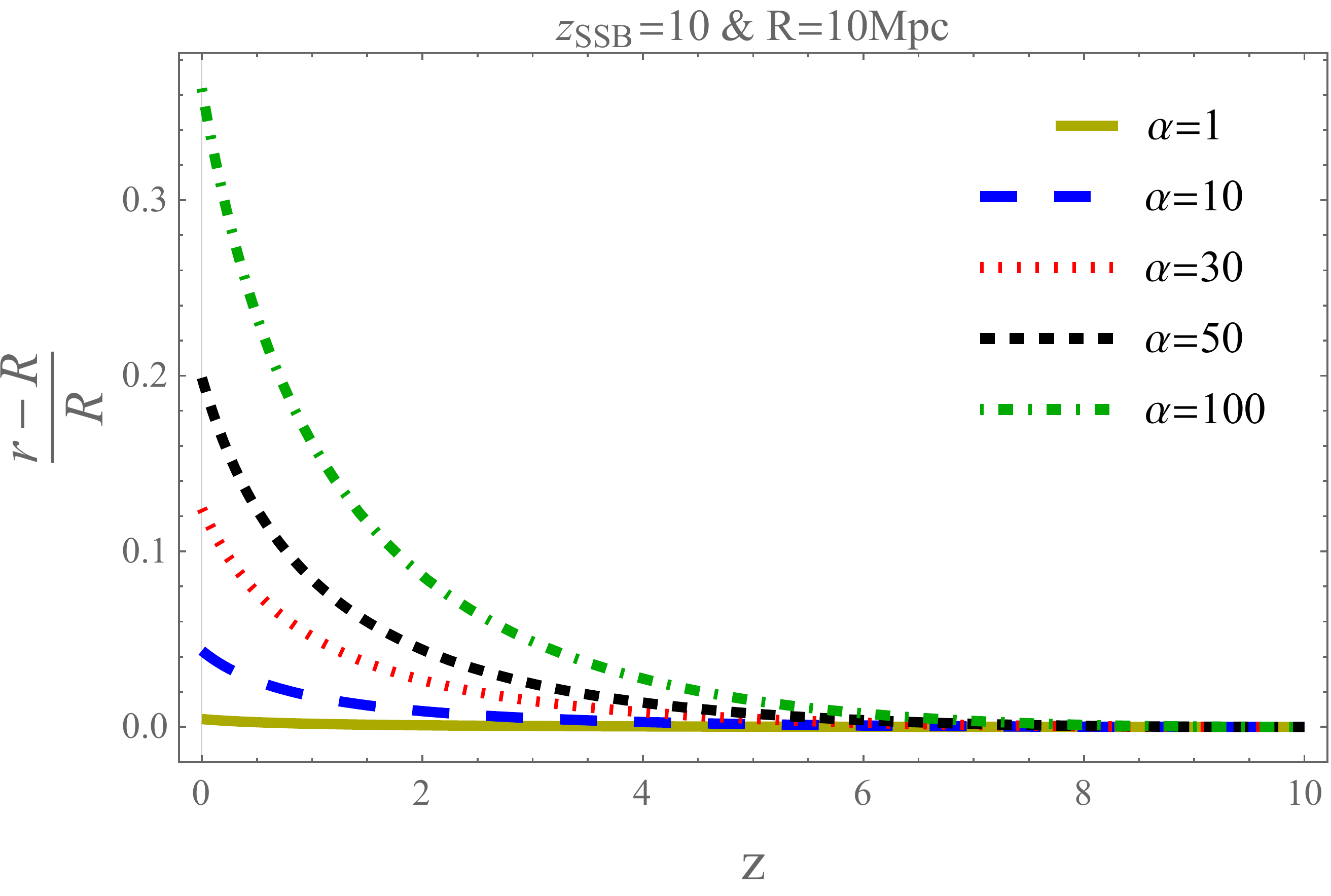}
	\caption{In this plot we have shown the affects of our free parameter (i.e. $\alpha$ which is introduced below relation (\ref{Veff-pot}) as $ \mu^2\,M^2 \sim \alpha\, M_{pl}^2\,H_{SSB}^2$) on the final results for a typical case of $R=10 Mpc$ and $z_{SSB}=10$.  The case $\alpha=1$ is the case which is considered in \cite{Llinares:2014zxa} which has a very (unobservable) tiny effect. But it is interesting that for one order of magnitude larger values of $\alpha$ we can get around $10\%$ increasing in the radius of voids.}
	\label{fig.R10.z10.Alpha}
\end{figure}
In figure \ref{fig.R10.z10.Alpha} we have shown the different values of the free parameter and it is evident that the slight change in $\alpha$ (one order of magnitude) can result in a larger voids (about $10\%$).

The interpretation of these results can be interesting. The voids are underdense regions which are surrounded by the (overdense) structures. It means when (symmetron's) DW pushes the structures away from the center of the voids, then the voids seem larger. This can be a smoking gun for the symmetron mechanism. More interestingly, this fact can be a solution for the void phenomenon. Since the DW's effect is a late time effect (via dark energy) then their effects on the nonlinear halo structures can be very small. This means while the voids are getting larger, the (number) distribution of the halos does not change. Consequently, we have bigger voids with the same number of halos in them i.e. the voids seem emptier in this model than in the standard model of cosmology. Note that our proposal is opposed to usual (cosmological) proposals for the void phenomenon. Usually, the proposals look for less number of halo structures inside the voids but we claim that voids are larger.

\section{Concluding remarks and future perspectives}
We studied the cosmological effects of DWs which are produced by the symmetron mechanisms. This is important as a test for these kinds of models beyond the standard probes e.g. the late time acceleration. The existence of these DWs can be consistent with the observational constraints on the parameters of the symmetron model. These DWs are placed in between the voids and structures and can affect their relative positions. We showed that there is an additional repulsive force on the matter (particles) outside the DW while the metric inside the DW is ineffective. This makes voids look larger than their predicted size in $\Lambda$CDM. In other words, the voids are emptier in comparison to the standard ($\Lambda$CDM) voids. It is a very interesting result since it can be a solution for the famous void phenomenon \cite{Peebles:2001nv}. Usually, to solve this phenomenon, models try to lessen the population of the structures inside the voids \cite{Peebles:2001nv}. But our proposal opens another path to look at this problem: larger voids make less dense voids. This way of looking at the void phenomenon has another very specific prediction which makes it distinguishable from the other scenarios. Since in our proposal there is no modification for the structure formation inside the voids then we expect the same distribution of the structures inside the voids as the standard scenario. However, since the voids are getting bigger these structures should seem to appear more at the center of the voids. This fact is in the direction of recent observations \cite{Tavasoli:2021reo}.

The idea which is explored in this work can be seen as a new way to look for the effects of modified gravity models in future surveys, especially in cosmological voids. In addition, late time DWs can have their fingerprints in gravitational waves if they could evaporate. On the theory side, the production of DWs in the symmetron model can be studied in more detail especially by asking about their stabilities (in this paper we assumed the DWs are stable since they are pinned to matter \cite{Llinares:2014zxa}). One other interesting way to pursue our idea is to check what happens if we do not assume the voids to be spherically symmetric.

\section*{Acknowledgments}
We would like to thank Shant Baghram, Levon Pogosian, and Saeed Tavasoli for their comments on our manuscript.

\section*{Appendix} \label{Ap-A}
To calculate the static spherical DW metric, as it was stated, normal unit vectors are needed. One finds $n_a$ for the outside and inside of the DW as
\begin{eqnarray}
\nonumber
	n_{\mu}|_+&=&\Big(0,\,-\big(1+\frac{F}{r}\big)^{-\frac{1}{2}},\,0,\,0\Big)\,,
	\\
	n_{\mu}|_-&=&\Big(0,\,-1,\,0,\,0\Big)\,.
\end{eqnarray}
In addition, intrinsic metric is assumed to have the following form
\begin{eqnarray}
\nonumber
	h_{ab}|_+&=&diag\Big(e^A\,\big(1+\frac{F}{R}\big),\,-R^2,\,-R^2\,\sin^2\theta\Big)\,,
	\\
	h_{ab}|_-&=&diag\Big(e^D,\,-R^2,\,-R^2\,\sin^2\theta\Big)\,,
\end{eqnarray}
therefore; the first condition yields (\ref{1st.SSDW}). Here we have found the ratio of $\frac{e^H}{e^G}$. To find $F$ we should obtain the second condition. In order to do so, we need to calculate (\ref{eq.gamma.K}), and (\ref{eq.gamma.h}). Extrinsic curvature for the static wall from the outside metric is
\begin{eqnarray}
\nonumber
	K_{tt}|_+&=&-\frac{F}{2R^2}\,e^{A}\,,
	\\
	K_{\theta\theta}|_+&=&R\,\sqrt{1+\frac{F}{R}}\,,
	\\
\nonumber
	K_{\phi\phi}|_+&=&R\sqrt{1+\frac{F}{R}}\,\sin^2\theta\,,
\end{eqnarray}
and from the inside metric, extrinsic curvature has the following form
\begin{eqnarray}
\nonumber
	K_{tt}|_-&=&0\,,
	\\
	K_{\theta\theta}|_-&=&R\,,
	\\
\nonumber
	K_{\phi\phi}|_-&=&R\,\sin^2\theta\,.
\end{eqnarray}
We obtain $\gamma_{ab}$ by substituting the above equations in (\ref{eq.gamma.K})
\begin{eqnarray}\label{eq.gamma.SDW.K}
	\nonumber
	\gamma_{tt}&=&-\frac{F}{2R^2}\,e^{A}\,,
	\\
	\gamma_{\theta\theta}&=&R\,\big(\sqrt{1+\frac{F}{R}}-1\big)\,,
	\\
	\nonumber
	\gamma_{\phi\phi}&=&R\,\sin^2\theta\,\big(\sqrt{1+\frac{F}{R}}-1\big)\,.
\end{eqnarray}
In addition (\ref{eq.gamma.h}) shows that
\begin{eqnarray}\label{eq.gamma.SDW.h}
\nonumber
	\gamma_{tt}&=&-4\,\pi\,G\sigma\,e^{D}\,,
	\\
	\gamma_{\theta\theta}&=&4\,\pi\,G\,\sigma\,R^2\,,
	\\
\nonumber
	\gamma_{\phi\phi}&=&4\,\pi\,G\sigma\,R^2\,\sin^2\theta\,.
\end{eqnarray}
We can obtaine the constant $F$ by comparing (\ref{eq.gamma.SDW.K}) and (\ref{eq.gamma.SDW.h}) which will lead us to (\ref{2nd.SSDW.thth}).

\end{document}